\documentclass[preprint,aps]{revtex4}
\usepackage{graphicx}
\usepackage{amsmath}
\usepackage{longtable}
\usepackage{dcolumn}
\newcommand{\be}{\begin{equation}}
\newcommand{\ee}{\end{equation}}
\newcommand{\ben}{\begin{eqnarray}}
\newcommand{\een}{\end{eqnarray}}
\newcommand{\ra}{\rangle}

\newcommand{\go}{\rightarrow}

\begin{document}

\title{Multi-Qubit Systems: Highly Entangled States and Entanglement
Distribution}
\author{A. Borras$^1$, A.R. Plastino$^{1,2,3}$\footnote{Corresponding Author:
arplastino@maple.up.ac.za}, J. Batle$^1$, C. Zander$^2$,  M.
Casas$^1$, and A. Plastino$^3$} \affiliation{ $^1$Departament de
F\'\i sica and IMEDEA-CSIC, Universitat de les Illes Balears,
07122 Palma de Mallorca, Spain \\
$^2$Physics Department,
University of Pretoria,
Pretoria 0002, South Africa \\
$^3$National University La Plata and Conicet,
Casilla de Correo 727, La Plata 1900, Argentina.}%

\date{\today}

\begin{abstract}
 A comparison is made of various searching procedures,
based upon different entanglement measures or entanglement
indicators, for highly entangled multi-qubits states. In particular,
our present results are compared with those recently reported by
Brown et al. [J. Phys. A: Math. Gen. 2005 {\bf 38} 1119]. The statistical
distribution of entanglement values for the aforementioned
multi-qubit systems is also explored.

\vskip 2cm

\noindent
keywords: Multi-Qubit Systems, Quantum Entanglement
\end{abstract}


\maketitle

\section{Introduction}

 Quantum entanglement \cite{BZ06} is nowadays regarded
as constituting one of (if not the) most basic features of quantum
mechanics \cite{N00,LPS98,BEZ00}. The increasing interest
generated by this subject within the research community
\cite{GLM06,GMM04,PSW06,GLM03a,GLM03b,BCPP05a,BCPP06,BPRST00,BCPP05b,
BSSB05,HS00,BH07,MW02,B03,WH05a,WH05b,CZ05,LS05,S04,CMB04,AM06,CHDB05,FFP06}
has been greatly stimulated by the discovery of novel quantum
information processes \cite{N00,LPS98,BEZ00} (such as quantum
teleportation and superdense coding) that may lead to important
practical developments. The technological relevance of quantum
entanglement is not limited to the information technologies, but is
also at the basis of other interesting applications, such as quantum
metrology \cite{GLM06}. Besides its remarkable technological impact,
current research in quantum entanglement is contributing to a deeper
understanding of various basic aspects of quantum physics, such as,
for instance, the foundations of quantum statistical mechanics
\cite{GMM04,PSW06}. The relationship between entanglement and the
dynamical evolution of multipartite quantum systems
\cite{GLM03a,GLM03b,BCPP05a,BCPP06} constitutes another interesting
example.

 Due to its great relevance, both from the fundamental
and from the practical points of view, it is imperative to explore
and characterize all aspects of the quantum entanglement of
multipartite quantum systems. A considerable amount of research has
recently been devoted to the study of multi-qubit entanglement
measures defined as the sum of bipartite entanglement measures over
all (or an appropriate family of) the possible bi-partitions of the
full system
\cite{BSSB05,HS00,MW02,B03,WH05a,WH05b,CZ05,LS05,S04,CMB04,AM06,CHDB05}
(see also \cite{FFP06} for another approach, also based on
bi-partitions, to multi-artite entanglement).
In particular, Brown et al. \cite{BSSB05} have performed a numerical
search of multi-qubit states exhibiting a high value of an
entanglement measure defined in the aforementioned way, based upon
the negativity of the system's bi-partitions. The purpose of the
present work is twofold. On the one hand, we numerically determine
the distribution of entanglement values (according to four different
measures of multi-qubit entanglement based upon bi-partitions) of
pure states of three, four, and five qubits, and its relationship with
important particular states, such as the $|GHZ\rangle$ state. On the
other hand, we report the result of running numerical searches of
multi-qubit states (up to 7 qubits) exhibiting high entanglement
according to the alluded to four measures. The results obtained using
each of these four measures are compared to each other, and also
compared to those reported by Brown et al. \cite{BSSB05}.

The paper is organized as follows. Some basic properties of the
entanglement measures used here are reviewed in Section II. Our
results concerning the distribution of multi-qubit entanglement
measures for systems of 3, 4, and 5 qubits are reported and
discussed in Section III. Our algorithm for the search of states of
high entanglement is presented in Section IV, and the main results
obtained are discussed and compared with those reported by Brown et
al. Finally, some conclusions are drawn in Section V.

\section{Pure State Multipartite Entanglement Measures Based on the
Degree of Mixedness of Subsystems}

  Research on the properties and applications
of multipartite entanglement measures has attracted considerable
attention in recent years
  \cite{BSSB05,HS00,MW02,B03,WH05a,WH05b,CZ05,LS05,S04,CMB04,AM06,CHDB05}.
 One of the first practical entanglement measures for $N$-qubit pure
states $|\phi \rangle$ to be proposed was the one introduced by
Meyer and Wallach \cite{MW02}. It was
 later pointed out by Brennen \cite{B03}
 that the measure advanced by Meyer and Wallach
 is equivalent to the average of all the single-qubit
 linear entropies,

 \be \label{globalenta}
 Q(|\phi \rangle) \, = \, 2
 \left(
 1 - \sum_{k=1}^N tr \rho_k^2
 \right).
 \ee

 \noindent
 where $\rho_k, \,\,\, k=1, \ldots N$, denotes the
 marginal density matrix describing the $k$th qubit
 of the system after tracing out the rest.
 This quantity, often referred to as ``global entanglement" (GE),
 describes the average entanglement of each qubit
 of the system with the remaining $(N\!-\!1)$-qubits.
 The GE measure is widely regarded as a legitimate,
 useful and practical $N$-qubit entanglement measure
 \cite{B03,WH05a,WH05b,CZ05,LS05}. This measure
 is invariant under local unitary transformations and
 non-increasing on average under local quantum operations
 and classical communication. That is to say, $Q$ is
 an entanglement monotone.
 Another interesting feature of this measure is that it
 can be determined without the need for full quantum state
 tomography \cite{B03}. This measure proved to be useful
 in the study of several problems related to multipartite entanglement,
 such as entanglement generation by nearly random operators \cite{WH05a}
 and by operators characterized by special matrix element
 distributions \cite{WH05b}, thermal entanglement in multi-qubit
 Heisenberg models \cite{CZ05}, and multipartite entanglement in
 one-dimensional time-dependent Ising models \cite{LS05}.
 Other entanglement measures, based upon the average
 values of the linear entropies associated with
 more general partitions of the $N$-qubit systems into two subsystems
 (that is, involving not only the partitions of the system into a $1$-qubit subsystem
 and an $(N\!-\!1)$-subsystem) have also been recently
 explored \cite{S04,CMB04,AM06}.
 In particular, Scott \cite{S04} studied various interesting
 aspects of the family of multiqubit entanglement measures
 given by

 \be \label{mglobalent}
 Q_m(|\phi \rangle) \, = \, \frac{2^m}{2^m-1}
 \left(
 1 - \frac{m! (N-m)!}{N!} \sum_s tr \rho_s^2
 \right), \,\,\,\,\, m=1,\ldots, [N/2],
 \ee

 \noindent
 where the sum is taken over all the subsystems $s$
 constituted by $m$ qubits, $\rho_s$ are the
 concomitant marginal density matrices,
 and $[x]$ is the integer part of $x$.
 The quantities $Q_m$ correspond to the average
 entanglement between subsystems consisting of
 $m$ qubits and the remaining $N-m$ qubits.
 The measures $Q_m$ have been applied to
 the study of quantum error correcting codes
 and to the analysis of the (multipartite)
 entangling power of quantum evolutions \cite{S04}.

 Another way of characterizing the global amount of entanglement
exhibited by an $N$-qubit state is provided by the sum of the
(bi-partite) entanglement measures associated with the
 $2^{N-1}-1$ possible bi-partitions of the $N$-qubits system
\cite{BSSB05}. These entanglement measures are given,
essentially, by the degree of mixedness of the marginal density
matrices associated with each bi-partition. These degrees of
mixedness can be, in turn, evaluated in several ways. For instance,
we can use the von Neumann entropy, the linear entropy, or a
Renyi entropy of index $q$. In what follows we are going to
consider the following ways of computing the degrees
of mixedness of the marginal density matrices $\rho_i$,

\begin{itemize}

\item The linear entropy $S_L = 1-Tr[\rho_i^2]$.

\item The von Neumann entropy $S_{VN} = -Tr[\rho_i \log_2 \rho_i] $.

\item The Renyi entropy with $q\go \infty$,  $S_{Re}^{q\go \infty} = - \ln \lambda_k^{max}$,
where $\lambda_k$ are the eigenvalues of the marginal density
matrix. This particular instance of the Renyi entropy constitutes the
case (within the Renyi family) that differs the most from the
von Neumann entropy \cite{BCPP02,BCPP03}.
\end{itemize}

\noindent
Besides these measures we are also going to consider the ``negativity" as a measure of
the amount of entanglement associated with a given bi-partition. The negativity
is given by

\be
{\rm Neg.} = \sum |\alpha_i|,
\ee

\noindent where $\alpha_i$ are the negative eigenvalues of the
partial transpose matrix associated with a given bi-partition. The
global, multipartite entanglement measures associated with the sum
(over all bi-partitions) of each of these four quantities are here
going to be denoted, respectively, by $E_L$, $E_{VN}$, $E_{R}$, and
$E_N$.

Upper bounds for the four entanglement measures $E_L$, $E_{VN}$,
$E_{R}$, and $E_N$ can be established by considering an
(hypothetical) $N$-qubits pure state such that all its
marginal density matrices are fully mixed. These bounds can
be seen in Table 1. Notice, however, that these bounds may
not be reachable. For instance, there is no four qubit state
reaching the alluded bound \cite{HS00}.

\vskip 1cm

\begin{table}[h]
\begin{center}
\begin{tabular}{|c||c|c|c|c|c|}
  \hline
  N & 3 & 4 & 5 & 6 & 7 \\
  \hline
  $E_{L, max}$ & 1.5 & 4.25 & 10 & 23 & 49.875 \\
  $E_{VN, max}$ & 3 & 10 & 25 & 66 & 154 \\
  $E_{Re, max}$ & 2.07944154 & 6.93147181 & 17.3286795 & 45.7477139 & 106.744666 \\
  $E_{Neg, max}$ & 1.5 & 6.5 & 17.5 & 60.5 & 157.5 \\
  \hline
\end{tabular}
\end{center}
\caption{Upper bounds for the entanglement measures
$E_L$, $E_{VN}$, $E_{R}$, and
$E_N$.}
\label{1}
\end{table}

\vskip 1cm


\section{\bf Distribution of Multiqubit Entanglement}

 In this section we determine numerically the distribution
 of entanglement values corresponding to pure states of
 multi-qubit systems randomly generated according to the Haar
 measure. In Figures 1, 2, and 3 we plot (for systems of 3, 4, and 5
 qubits respectively) the probability densities $P$ of finding
 multi-qubit states with given values of the entanglement measures
 $E_L$, $E_{VN}$, $E_R$, and $E_N$. In these Figures
 we also show vertical lines corresponding to the entanglement
 values of important particular states, such as the $N$-qubit
 GHZ state,

 \be
|GHZ\rangle \, = \, \frac{1}{\sqrt{2}}(
|0\ldots 0\rangle + |1\ldots 1 \rangle),
 \ee

\noindent the states of high entanglement $BSSB4$ and $BSSB5$ (of
four and five qubits, respectively) discovered numerically by Brown
et al. \cite{BSSB05}, and the four qubit state $HS$, that has
 been conjectured to maximize the entanglement of four-qubit
states \cite{HS00} (when measuring entanglement using the
sum of the marginal
von Neumann entropies associated with all bi-partitions).
The $HS$ state has recently been shown to constitute
a local maximum of the $E_{VN}$ entanglement measure for four-qubits
states \cite{BH07}.

\vskip 5cm

\begin{figure}[h]
\begin{center}
\hskip 7cm{\includegraphics[scale=0.5,angle=270]{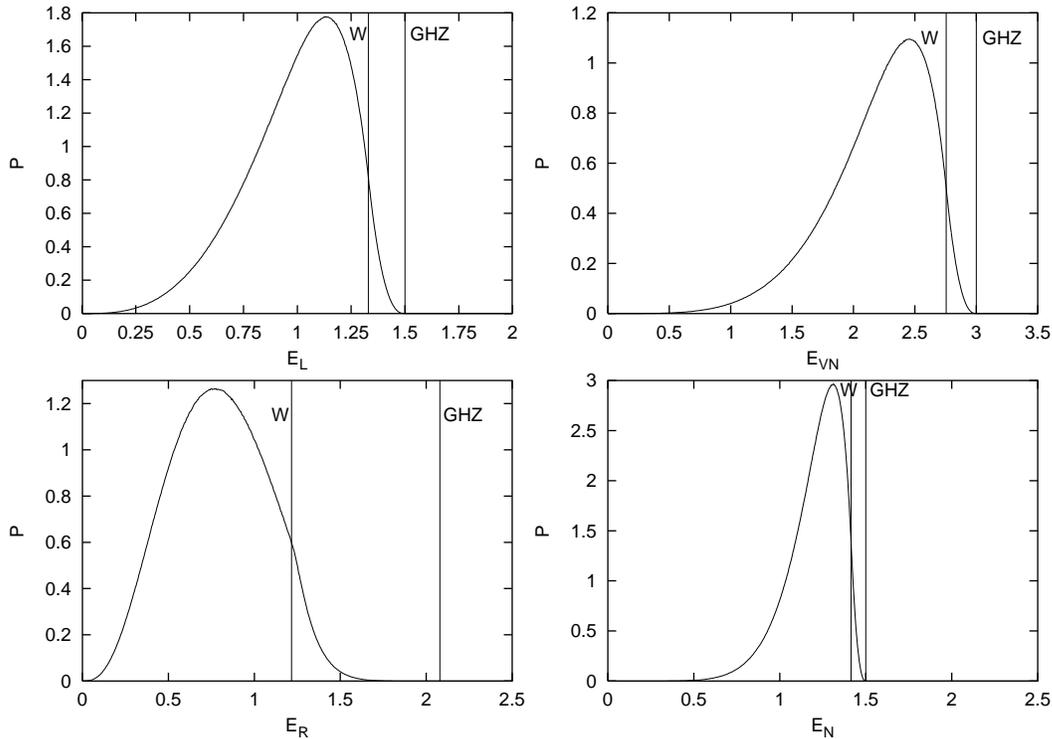}}
\vskip 1cm \center\caption{Entanglement distributions for 3 qubits
states. All depicted quantities are dimensionless.} \label{fig1}
\end{center}
\end{figure}

 A particularly interesting aspect of Figures 1, 2, and 3 is the
status (as far as the present multi-qubit entanglement measures
are concerned) of the state GHZ with respect to the bulk of
the states of the multi-qubit system.

 For three qubits systems, the $|GHZ\rangle$ state has all its
 single-qubit marginal density matrices complete mixed and,
 consequently, constitutes the state of maximum entanglement
 according to the measures $E_{VN}$, $E_L$, $E_N$, and $E_R$.
 On the other hand, the state

 \be
|W\rangle \, = \, \frac{1}{\sqrt{3}} \Bigl( |100\rangle +
|010\rangle + |001\rangle \Bigr),
 \ee

\noindent  according to those same measures, exhibits considerably
less entanglement than $|GHZ\rangle$. However, as can clearly be
appreciated in Figure 1, the $W$ state is still within the most
entangled three-qubit pure states. The $W$ state is clearly more
entangled than the ``typical" pure state of three qubits.

We have seen that, in the case of three-qubits the four measures
$E_{VN}$, $E_L$, $E_N$, and $E_R$ lead to qualitatively similar
conclusions in connection with the entanglement of the states $GHZ$
and $W$ as compared with the entanglement exhibited by typical
(pure) states. On the contrary, when four-qubit states are
considered, each of the aforementioned entanglement measures yields
different results. According to $E_R$, the state $|GHZ\rangle$ still
has an amount of entanglement well above most pure states.
According to $E_L$, the state $|GHZ\rangle$ has an entanglement a
little above typical. According to $E_{VN}$, $|GHZ\rangle$ can be
said to be (in terms of its entanglement value) still
``within the bulk of pure states", but with an
amount of entanglement clearly below typical. Finally, according to
$E_{N}$, the $|GHZ\rangle$ state exhibits less entanglement than
most pure states of four qubits. It is also interesting to
notice that the state $HS$ exhibits more entanglement
than $BSSB4$ when using the measures $E_L$, $E_{VN}$, or $E_N$.
On the contrary, $BSSB4$ has a larger value of
$E_R$ than $HS$.

\vskip 5cm

\begin{figure}[h]
\begin{center}
\hskip 7cm{\includegraphics[scale=0.5,angle=270]{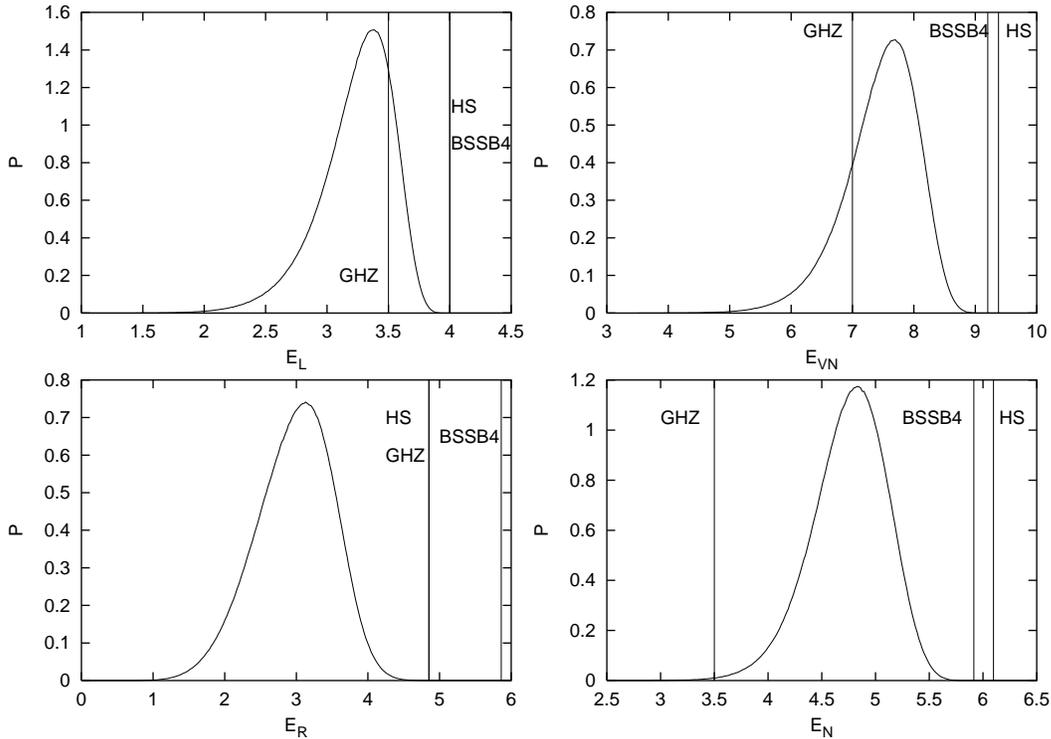}}
\vskip 1cm \caption{Entanglement distributions for 4 qubits states.
All depicted quantities are dimensionless.}
\end{center}
\label{fig2}
\end{figure}

For five-qubit states, the
$|GHZ\rangle$ state has less entanglement than most pure states when
the entanglement is measured using $E_L$, $E_{VN}$, or $E_N$.
Curiously enough, according to $E_R$ the $|GHZ\rangle$ still ranks
as a five-qubit state of rather large entanglement.

\section{\bf Search for Multi-Qubit States of High Entanglement}

\subsection {Searching  Algorithm}

 In the present paper we are going to restrict our search
of multi-qubit states of high entanglement to {\it pure states}.
In this respect our approach is a little different from that of
Brown et al. \cite{BSSB05}, who considered a search process
within the complete space of possible states (that is, with any
degree of mixedness). The kind of search studied by Brown et al. is
certainly of interest and may shed some light on the structure of the
``entanglement landscape" of the full state space. However,
it is reasonable to expect the states of maximum entanglement
to be pure. Consequently, as far as the search of states of maximum
entanglement is concerned, it seems that limiting the search to
pure states is not going to reduce its efficiency. The results
reported here fully confirm this expectation.

\vskip 5cm

\begin{figure}[h]
\begin{center}
\hskip 7cm{\includegraphics[scale=0.5,angle=270]{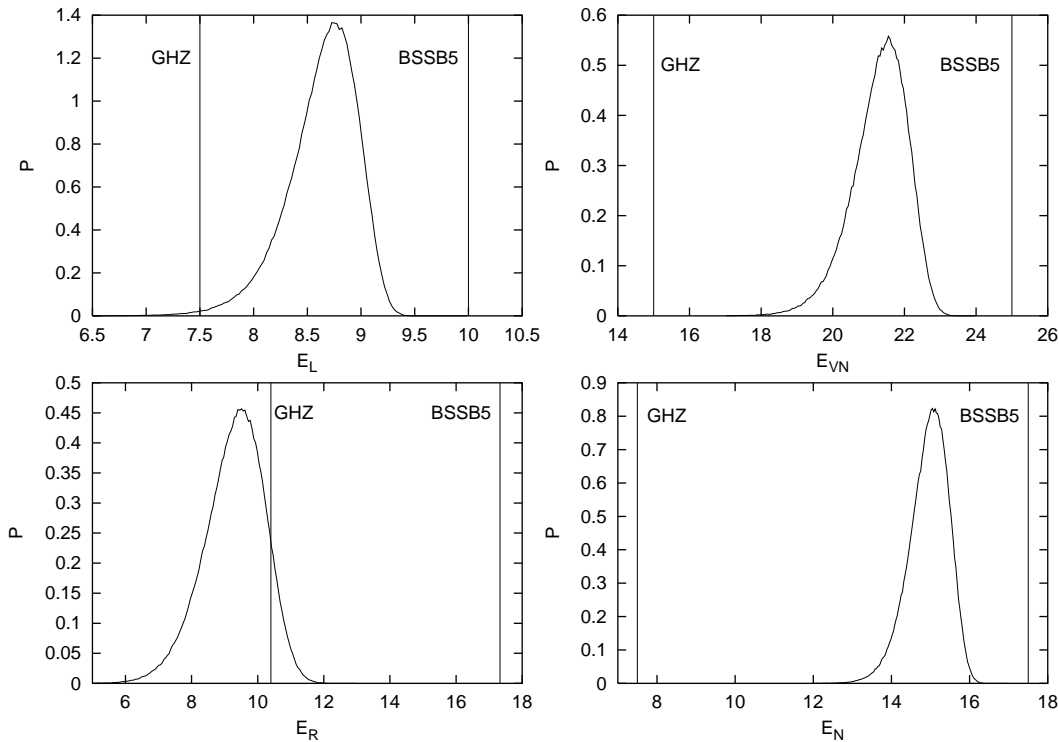}}
\vskip 1cm \caption{Entanglement distributions for 5 qubits states.
All depicted quantities are dimensionless.}
\end{center}
\label{fig3}
\end{figure}

A general pure state of an $N$-qubit system can be represented
as

\be
|\Psi \rangle \, = \, \sum_{k=1}^{2^N} (a_k + i b_k) |k\rangle,
\ee

\noindent where $|k\rangle$, $\,\,(k=1,\ldots, 2^N)$ represents the
states of the computational basis (that is, the $2^N$ states
$|00,\ldots, 0\rangle$, $|10,\ldots, 0\rangle$, $\ldots,$
$|11,\ldots, 1\rangle$). We start our search process with the
initial state $|000...0\rangle$. In other words, the initial
parameters characterizing the state are $a_1=1$, and all the rest of
the $a_i$'s and $b_i$'s are equal to zero. This initial state is
fully factorizable and can thus be regarded as being ``very distant"
from states of high entanglement. Starting with an
arbitrary, random initial pure state does not alter the results of
the search process. Now, at each step of the search process a new,
tentative state is generated according to the following procedure. A
random quantity $\Delta$ (uniformly chosen from an interval
$(-\Delta_{max}, \Delta_{max})$) is added to each $a_i$ and $b_i$ (a
different, independent $\Delta$ is generated for each parameter).
The new state generated in this way is then normalized to $1$ and
its entanglement measure is computed. If the entanglement of the new
state is larger than the entanglement of the previous state the new
state is kept, replacing the previous one. Otherwise, the new state
is rejected and a new, tentative state is generated. In order to
ensure the convergence of this algorithm to a state of high
entanglement, the following two rules are also implemented,

\begin{itemize}

\item{ If 500 consecutive tentative new states are rejected,
the interval for the random quantity $\Delta $ is changed according
to $\Delta_{max} \rightarrow \frac{\Delta_{max}}{2}$ (as the initial
value for $\Delta_{max}$ we take $\Delta_{max}^{init}=0.1$).}

\item{ When a value $\Delta_{max} \le 1 \cdot 10^{-8}$ is reached
the search program halts.}

\end{itemize}

\subsection {Results Yielded by the Searching Algorithm}

The maximum entanglement values obtained from the searching
algorithm are listed in Table 2. It must be stressed that the
maximum values associated with different measures do not necessarily
correspond to the same state. The states obtained when maximizing
one particular measure do not exhibit, in general, a maximum value
of the other measures. The results obtained by us after running the
search algorithm several times (considering the entanglement
measures $E_L$, $E_{VN}$, $E_R$, and $E_N$) can be summarized as
follows,

\begin{itemize}
\item{Among the four measures considered here, $E_L$ is computationally
the easiest and quickest to evaluate. The algorithm runs faster when
maximizing this measure than when maximizing any of the other three.
However, in the case of four-qubits most states that maximize $E_L$
do not maximize the other measures. There are {\it many} different
four qubit states that exhibit the observed maximum value $E_L=4$.
Few of these states exhibit also the maximum value of the other
entanglement measures (for instance, the value $E_{VN}=9.37734$).}

\item{The measure $E_{VN}$ is computationally more expensive than $E_L$.
The states obtained maximizing $E_{VN}$ also maximize $E_L$ and
$E_N$. In other words, all the states that we have found that realize
the observed maximum value of $E_{VN}$ realize as well the observed
maxima of $E_L$ and $E_N$. On the contrary. for four qubits there are
many states exhibiting the observed maximum value of $E_L$ that do not
reach the observed maxium value of $E_{VN}$.}

\item{The measure $E_{R}$ seems to be the ``worst" of the four.
States that maximize $E_{R}$ do not, in general, maximize the other
measures. And, conversely, states maximizing any of the other
measures do not in general maximize $E_{R}$.}

\item{$E_N$ is, by far, computationally the most expensive of the
measures considered here. The states maximizing this measure also
maximize $E_L$ and $E_{VN}$. In this case the situation is similar
to the already mentioned one corresponding to the measure $E_{VN}$.}
\end{itemize}





\vskip 1cm

\begin{table}[h]
\begin{center}
\begin{tabular}{|c|c|c|c|c|c|}
  \hline
    & 3 qubits & 4 qubits & 5 qubits & 6 qubits & 7 qubits\\ \hline
  $E_L$ & 1.500000 & 4.00000 & 10.000000 & 23.000000 & 49.573765 \\
  $ E_{VN} $ & 3.000000 & 9.37734 & 25.000000 & 66.000000 & 152.620140 \\
  $ E_{R} $ & 2.079441 & 5.99547 & 17.328678 & 45.747705 & 91.651820  \\
  $ E_N $ & 1.500000 & 6.09807 & 17.500000 & 60.500000 & 155.812856 \\
  \hline
\end{tabular}
\end{center}
\caption{Numerically obtained maximum values for the
entanglement measures $E_L$, $E_{VN}$, $E_{R}$, and
$E_N$.}
\label{2}
\end{table}

\vspace{1cm}

 The numerical values reported in the above Table are the
result of several search experiments that can be summarized
as follows. In the case of three qubits the numerical
optimization of any of the aforementioned measures leads
to the same state, the $|GHZ\rangle$ state, and to the
concomitant maxima of the entanglement measures.
For four qubits, the search for states optimizing $E_{VN}$ yields a final
state (equivalent to the $|HS\ra$ state) that also maximizes all the
entanglement measures considered excepting $E_{R}$ (here, by ``equivalent
to the  $|HS\ra$ state" we mean that all the marginal density matrices
of the alluded state exhibit the same entropic values
as those exhibited by the corresponding marginal density matrices
of $|HS\ra$, and also that the alluded state has, for all bipartitions,
the same negativities as  $|HS\ra$). The maximum value of $E_{R}$
reported in Table 2 is generated by search experiments
maximizing this entanglement measure. The explicit expression of the
corresponding four qubits state is given in
the Apendix. The four qubit state obtained when
searching for the maximum value of  $E_N$
is equivalent  to the one
obtained when maximizing $E_{VN}$. When conducting search
experiments for four qubit states maximizing  $E_L$
we obtain, in most cases, states that do not reach
the observed maxima of the rest of the measures. These
states are not, in general, equivalent to each other.
In point of fact, a different state is obtained in each
run of the algorithm optimizing $E_L$.

The five qubits case is similar to the three qubits one.
The numerical search of five qubit states optimizing any
of the aforementioned entanglement measures leads to
states that exhibit the observed maxima of all these
measures (which are reported in Table 2). In other words,
if one runs a search algorithm based upon any one of
these measures, one obtains a state that exhibits
all the maximum entanglement values reported in Table 2.
These values are the ones corresponding to the five
qubits state (\ref{cincoqu}).

For six qubits, the search experiments based on the
maximization of either $E_L$ or  $E_{VN}$ lead to
final states exhibiting the same values of the four
entanglement measures, which are reported in Table 2. The search
algorithm based upon the optimization of $E_{R}$
yields states with lower values of the four measures
than those shown in Table 2. For six qubits the
search algorithm corresponding to $E_N$ is too
slow and we were not able to reach the optimal
state.

Finally, in the case of seven qubits the values
reported on Table 2 were evaluated on the state
found when numerically optimizing $E_{VN}$ (this
state is explicitly given in the Appendix). When
running numerical searches for seven qubit states
optimizing other measures we did not find states
with entanglement values higher than those
evaluated upon the state obtained when optimizing
$E_{VN}$.

Let us now discuss in more detail the numerically
found states of high entanglement.

\subsubsection{Four Qubits}

In the case of four-qubit systems, the extremalization processes
based upon either of the measures $E_{VN}$ or $E_N$ always lead to
states having the same entanglement values as those exhibited by the
$HS$ state discovered by Higuchi and Sudbery \cite{HS00}, which is
given by

\be |HS\rangle \, = \, \frac{1}{\sqrt{6}} \Bigl[ |1100\rangle+
|0011\rangle + \omega \Bigl(|1001\rangle + |0110\rangle \Bigr) +
\omega^2 \Bigl(|1010\rangle + |0101\rangle\Bigr) \Bigr], \ee

\noindent with $\omega= -\frac{1}{2}+\frac{\sqrt{3}}{2}$. We
repeated the search process starting with different, random initial
conditions and always found states with entanglement values
corresponding to the $HS$ state. This constitutes convincing numerical
evidence that the $HS$ state is, at least, a local maximum of {\it both}
the $E_{VN}$ and the $E_N$ measures. In fact, it was recently
proven by Brierley and Higuchi that the $HS$ state is indeed
a local maximum for $E_{VN}$ \cite{BH07}. Higuchi and Sudbery \cite{HS00}
have provided analytical arguments supporting the conjecture that
the $HS$ state is also a {\it global maximum} for $E_{VN}$, but this
conjecture has not been proven yet. These authors have also proved
that there is no pure state of four qubits such that all its
two-qubit marginal density matrices are completely mixed
\cite{HS00}. It is interesting that Brown et al. \cite{BSSB05}, when
performing a search process similar (but not identical) to the one
considered here, obtained instead of the $HS$ state always
a state (which we here call $BSSB4$) exhibiting  values of
$E_{VN}$ and $E_N$ smaller than those exhibited by $HS$. Besides
some intrinsic differences in the algorithm itself, there is the
fact that the main results reported here were computed starting the
search process with a pure state, while Brown et al. started their
search with a mixed state. It is also worthwhile mentioning that we
performed the searches using a FORTRAN program, while Brown et al.
employed a MAPLE program. When running a search algorithm maximizing
the $E_L$ measure, we obtained several different final states, some
of them exhibiting values of $E_{VN}$ larger than the value
corresponding to the state $BSSB4$. All these findings suggest that,
perhaps, the state $BSSB4$ has no special significance (although it
certainly is a highly entangled four-qubits state). Its
appearance when running the searching scheme developed by Brown et
al. seems to be just an accident due to some special features of that
algorithm.

We must mention that we also ran a search
algorithm (written in the computer language MATHEMATICA) similar to
that of Brown et al. (and different from the one discussed in most
of the present paper), obtaining the same results
as Brown et al. did (that is, the algorithm converged to a
state with entanglement values corresponding to $BSSB4$).
 On the other hand, when running an algorithm (written in
MATHEMATICA) exhibiting the same basic structure of our FORTRAN
program we get the same results as those obtained with the
FORTRAN code. The main difference between our algorithm (either in
the FORTRAN or the MATHEMATICA versions) and the one used by
Brown et al. (when particularized to pure states) is the following.
When generating new random trial states (in the ``pure state version"
of Brown et al. algorithm) one choses a random
coefficient of the previous state, multiply the corresponding
real and imaginary parts by positive random numbers, and re-normalize
the state. On the other hand, in our algorithm (see sub-section IV A)
{\it we add random
numbers (that may be positive or negative) to the real and imaginary
parts of the state's coefficients} (and then re-normalize the state).
The results of various numerical
experiments done by us suggest that this difference on the implementation
of the searching algorithm accounts for the different results
obtained for highly entangled four qubits states.

\subsubsection{Five Qubits}

When running our search scheme for states of five qubits, we always
obtain states exhibiting the same entanglement values as the state
obtained by Brown et al. \cite{BSSB05},

\be \label{cincoqu}
|BSSB5\rangle \, = \, \frac{1}{2} \Bigl[ |100\rangle
|\Phi_{-}\rangle +|010\rangle |\Psi_{-}\rangle +
|100\rangle|\Phi_{+}\rangle + |111\rangle|\Psi_{+}\rangle\Bigr]
 \ee

\noindent where $\Psi_{\pm}=|00\rangle \pm |11\rangle$ and
$\Phi_{\pm}=|01\rangle \pm |10\rangle$. This state has all its
marginal density matrices (for 1 and 2 qubits) completely mixed.

\subsubsection{Six Qubits}

In the case of six qubits, our algorithm converges to highly
entangled states exhibiting all the marginal density matrices for
states of 1, 2, 3 qubits completely mixed. In particular, we
discovered the new state of high entanglement,

 \begin{eqnarray}
\Psi_{6qb} = \frac{1}{\sqrt{32}} \, \Big[ \, |000000> + |111111> +
|000011> + |111100> + |000101> + |111010> \nonumber \\ + |000110> +
|111001> + |001001> + |110110> + |001111> + |110000> \nonumber \\ +
|010001> + |101110> + |010010> + |101101> + |011000> + |100111> \nonumber \\
+ |011101> + |100010> - \big( \, |001010> + |110101> + |001100> +
|110011> \nonumber \\ + |010100> + |101011> + |010111> + |101000> +
|011011> + |100100> \nonumber \\ + |011110> + |100001> \big) \, \Big]
\end{eqnarray}

\noindent which, to the best of our knowledge, has not yet been
reported in the literature. This state has a rather simple
structure, with all its coefficients (when expanded in the
computational basis) equal to $0$ or $\pm 1$ (the same situation
occurs for maximally entangled states of 2, 3, 4, and 5 qubits).

\subsubsection { 7 qubits }

 When we ran the search program for seven-qubit states of high
entanglement we found states with the following features. They all
have completely mixed single qubit marginal density matrices.
However, these states do not exhibit completely mixed two-qubit and
three-qubit marginal density matrices (in this sense, the present
situation seems to have some similarities with the four-qubit case).

The high entanglement states of seven qubits that we found
are characterized by two-qubits marginal density matrices exhibiting
the following entropic values

\be 1 - Tr(\rho_i^2) =
0.7445111988 \ee \be S_{VN} (\rho_i) = 1.9841042 \ee \be
S_{Re}^{q\go\infty} (\rho_i) = 1.248122309. \ee


\noindent
The three-qubit marginal density matrices of these seven-qubit
states have,

\be 1 - Tr(\rho_i^2) =
0.86209018886 \ee \be S_{VN} (\rho_i) = 2.93739788 \ee \be
S_{Re}^{q\go\infty} (\rho_i) = 1.4712659418. \ee


 When running our program (maximizing either
 $E_{VN}$ or $E_N$) for five-qubit or six-qubit states, {\it the search
 process always leads to a state whose marginal density matrices of 1,2,
 and (in the six-qubit case) 3 qubits are completely mixed}. On the
 contrary, this never happens when running our algorithm for seven-qubits
 states. The marginal density matrices of 1 qubit subsystems turn
 out to be maximally mixed, but not the marginal density matrices
 corresponding to subsystems consisting of 2 or 3 qubits. Moreover,
 all the runs of the algorithm for seven-qubits states yielded
 states with the same entropic values for the marginal statistical
 operators. This suggests that the case of seven qubits may have
 some similarities with the case of four qubits. In other words,
 our results constitute numerical evidence supporting the

 \noindent
  {\bf Conjecture 1:} {\it There is no pure state of seven qubits whose
 marginal density matrices for subsystems of 1, 2, or 3 qubits are all
 completely mixed}.

 \subsubsection{The Single-Qubit Reduced States Conjecture}

It was conjectured by Brown et al. \cite{BSSB05} that multi-qubit
states of maximum entanglement always have all their single-qubit
marginal density matrices completely mixed. The results obtained by
us when running the search algorithm maximizing the $E_{VN}$ and
$E_N$ measures are consistent with the aforementioned conjecture.
All the states yielded by the searching algorithm (up to systems of
seven qubits) have maximally mixed single qubit marginal density
matrices. Moreover, in the case of 5 qubits all the states obtained
also exhibited maximally mixed two-qubits marginal density matrices.
In the case of 6 qubits, all the states obtained had completely
mixed marginal density matrices of one, two, and three qubits.

\section {Conclusions}

 In the present effort we have investigated some aspects of
the entanglement properties of multi-qubit systems. We have
considered global, multi-qubit entanglement measures based upon the
idea of considering all the possible bi-partitions of the system.
For each bi-partition we computed a bi-partite entanglement measure
(such as the von Neumann entropy of the marginal density matrix
associated with the subsystem with a Hilbert space of lower
dimensionality) and then summed the measures associated with all the
bi-partitions. This approach has been widely used in the recent
literature. In order to evaluate the bi-partite contributions we
considered four different quantities: the von Neumann, linear, and
Renyi (with $q\go \infty$) entropies, and the negativity. Consequently,
we have considered four entanglement measures.

 We determined numerically, for the aforementioned four measures,
the distributions of entanglement values in the Hilbert spaces of
pure states of three, four, and five qubits. This allowed us to
determine, for instance, the entanglement status of special states
(such us the $|GHZ\rangle$ state) with respect to the bulk of the
state space.

 We also determined, for systems of four, five, six and seven qubits,
states of high entanglement using a search scheme akin, but not
identical to, the one recently advanced by Brown et al.
\cite{BSSB05}. These authors performed the search process using an
entanglement measure based on the negativity. We investigated the
behavior of the search processes based on four different measures:
the negativity, and the von Neumann, linear, and Renyi (with
$q \go \infty$) entropies of the marginal density matrices associated
with a bi-partition. The results obtained by us have some
interesting features when compared with those reported by Brown et
all \cite{BSSB05}. First of all, we found that a search algorithm
based on the von Neumann entropy is as successful as one based upon
negativity. However, the von Neumann entropy is (in general)
considerably less expensive to compute than the negativity.
Consequently, when initializing the search process with a pure
state, it is better to use the von Neumann entropy.

 In the case of states of four qubits Brown et al. reported that
 their search algorithm always converged (up to local unitary
 transformations) to a state (here called the $BSSB4$ state)
 exhibiting less entanglement than the $HS$ state. On the contrary,
 our algorithm always converged to states exhibiting the same
 entanglement measures as those characterizing the $HS$ state.
 Our results thus provide further support to the conjecture
 advanced by Higuchi and Sudbery \cite{HS00} that the $HS$
 state corresponds to a global entanglement maximum for
 four-qubits states. Another interesting finding, going
 beyond the results of Brown et al. is a particular state of six qubits
 (discovered using our search algorithm) that has all its marginal
 density matrices of 1, 2, and 3 qubits completely mixed. It is interesting
 that (in the computational basis) all the coefficients characterizing this
 state are (up to a global normalization constant) equal to $0$ or $\pm 1$.

 Finally, on the basis of the numerical evidence obtained by us
 when running our search algorithm for highly entangled states
 of seven qubits, we make the conjecture that there is no pure
 state of seven qubits whose marginal density matrices for subsystems
 of 1, 2, or 3 qubits are all completely mixed.

\vskip 3mm {\bf Acknowledgements} \vskip 3mm
This work was partially supported by the
MEC grant FIS2005-02796 (Spain) and FEDER (EU) and
by CONICET (Argentine Agency). $\,$
The financial assistance of the National Research Foundation
(NRF; South African Agency) towards this research is hereby
acknowledged. Opinions expressed and conclusions arrived at,
are those of the authors and are not necessarily to be
attributed to the NRF. $\,$
A. Borr\'as acknowledges support from
the FPU grant AP-2004-2962 (MEC-Spain).

\newpage

\section{Appendix}
In this appendix we present the explicit expressions for some
of the states that we have introduced in the previous sections.
To give the expression of a state $| \Psi \ra$ we list the
values of the coefficients $C_i$ appearing in the expansion
$| \Psi \ra = \sum C_i |i\ra$ of the alluded
state in the computational basis $\{|i\ra \}$.

\begin{center}
\begin{longtable}{|r|c|}
\caption[State for 4 qubits]{Coefficients for the 4 qubit
state maximizing the entanglement measure based on the Renyi
entropy. This state doesn't maximize any other entanglement measure.}
\\
\hline \multicolumn{1}{|c|}{\textbf{$ \ \ \ \ \ i \ \ \ \ \ $}} &
\multicolumn{1}{c|}{\ \ \ \ \ \ \ \ \ \ \ \ \ \ \ \ \ \ \ \ \ \ \
\ \ \ \ \ \ \ \ \ \ \ \ \ \ \ \ \ \ \ \ \ \ \ \ \ \textbf{$ C_i $}
\ \ \ \ \ \ \ \ \ \ \ \ \ \ \ \ \ \ \ \ \ \ \ \ \ \ \ \ \ \ \ \ \
\ \ \ \ \ \ \ \ \ \ \ \ \ \ \ }  \\ \hline
\endfirsthead

\multicolumn{2}{c}%
{{\bfseries \tablename\ \thetable{} -- continued from previous page}} \\
\hline \multicolumn{1}{|c|}{\textbf{$ \ \ \ \ \ i \ \ \ \ \ $}} &
\multicolumn{1}{c|}{\ \ \ \ \ \ \ \ \ \ \ \ \ \ \ \ \ \ \ \ \ \ \
\ \ \ \ \ \ \ \ \ \ \ \ \ \ \ \ \ \ \ \ \ \ \ \ \ \textbf{$ C_i $}
\ \ \ \ \ \ \ \ \ \ \ \ \ \ \ \ \ \ \ \ \ \ \ \ \ \ \ \ \ \ \ \ \
\ \ \ \ \ \ \ \ \ \ \ \ \ \ \ }  \\ \hline
\endhead

\hline \multicolumn{2}{|r|}{{Continued on next page}} \\ \hline
\endfoot

\hline
\endlastfoot

0 & (0.337140676904686,0.174693405076796)             \\
  1 & (3.860442882346969E-002,6.837682483380016E-002)   \\
  2 & (5.962390590615981E-002,0.130590439038055)        \\
  3 & (3.780903708091862E-002,0.283134470502957)        \\
  4 & (0.128308013031141,0.160044519815334)             \\
  5 & (-4.976588113149925E-002,-0.156794899004251)      \\
  6 & (0.150158286657780,-0.269632673631216)            \\
  7 & (-0.284880375838561,4.364132887880368E-002)       \\
  8 & (-0.291078649973983,-0.122251701129522)           \\
  9 & (8.597952221078008E-002,-0.132269103402589)       \\
  10 &(-0.184679774192993,-3.521179357675151E-002)      \\
  11 &(-7.859668707973404E-002,0.285246180204626)       \\
  12 &(-3.120148147808102E-002,3.966923168894761E-002)  \\
  13 &(-0.352475250278756,-0.170787520712258)           \\
  14 &(2.666941273479068E-002,-0.244143026082971)       \\
  15 &(0.176830325000684,-7.078443862056820E-002)       \\
 \hline
\end{longtable}
\end{center}

\vspace{1cm}

\begin{center}
\begin{longtable}{|r|c|}
\caption[State for 7 qubits]{Coefficients for the 7
qubit state maximizing the von Neumann entropy based
entanglement measure. It also maximizes the rest of
the entanglement measures used along this paper.}
\label{7qbstateVN} \\
\hline \multicolumn{1}{|c|}{\textbf{$ \ \ \ \ \ i \ \ \ \ \ $}} &
\multicolumn{1}{c|}{\ \ \ \ \ \ \ \ \ \ \ \ \ \ \ \ \ \ \ \ \ \
\ \ \ \ \ \ \ \ \ \ \ \ \ \ \ \ \ \ \ \ \ \ \ \ \ \ \textbf{$ C_i $}
\ \ \ \ \ \ \ \ \ \ \ \ \ \ \ \ \ \ \ \ \ \ \ \ \ \ \ \ \ \ \ \ \
\ \ \ \ \ \ \ \ \ \ \ \ \ \ \ }  \\ \hline
\endfirsthead

\multicolumn{2}{c}%
{{\bfseries \tablename\ \thetable{} -- continued from previous page}} \\
\hline \multicolumn{1}{|c|}{\textbf{$ \ \ \ \ \ i \ \ \ \ \ $}} &
\multicolumn{1}{c|}{\ \ \ \ \ \ \ \ \ \ \ \ \ \ \ \ \ \ \ \ \ \
\ \ \ \ \ \ \ \ \ \ \ \ \ \ \ \ \ \ \ \ \ \ \ \ \ \ \textbf{$ C_i $}\
\ \ \ \ \ \ \ \ \ \ \ \ \ \ \ \ \ \ \ \ \ \ \ \ \ \ \ \ \ \ \ \ \ \ \
\ \ \ \ \ \ \ \ \ \ \ \ }  \\ \hline
\endhead
\hline \multicolumn{2}{|r|}{{Continued on next page}} \\ \hline
\endfoot
\hline
\endlastfoot

  0 & (1.992268895612789E-002,-2.048153299374923E-002)  \\
  1 & (5.733894334752334E-002,4.973994982020743E-003) \\
  2 & (-4.620635677624599E-002,9.889188153518157E-002) \\
  3 & (0.114773068934711,7.803541807299509E-002) \\
  4 & (9.358057357464943E-003,8.773453313011471E-002) \\
  5 & (-4.517771306482277E-002,7.317172187520525E-002) \\
  6 & (7.148596275123295E-002,-6.486415242189469E-002) \\
  7 & (8.095549161110917E-002,6.281081599967211E-002) \\
  8 & (-0.110934833126726,-6.540485101339541E-002) \\
  9 & (4.243711009834195E-002,0.111608997849607) \\
  10 & (-5.324057236738998E-002,-1.064133868681598E-002) \\
  11 & (-3.199776618312627E-002,1.480812105331856E-002) \\
  12 & (-3.484102446829535E-002,6.505443761669717E-002) \\
  13 & (6.659331311799828E-002,2.520078454850319E-002) \\
  14 & (2.127875261481843E-002,-8.620489194999095E-003) \\
  15 & (3.763178050938378E-002,-3.257033322657695E-002) \\
  16 & (-9.639113945809372E-002,-8.706895542690339E-002) \\
  17 & (7.213494811044056E-002,1.637328607897790E-002) \\
  18 & (3.347204156200859E-003,-4.540542385699349E-002) \\
  19 & (5.235538552827945E-002,-5.539353156272388E-002) \\
  20 & (-5.734329608600269E-002,-3.334326701130044E-002) \\
  21 & (-2.042578560682204E-002,-0.106743556238253) \\
  22 & (-5.987692237756689E-002,-5.035304599306584E-002) \\
  23 & (3.304680530465200E-002,9.449073856519782E-002) \\
  24 & (2.843182057391498E-002,-2.453794986457519E-002) \\
  25 & (-1.316539219004622E-002,-4.912228258199161E-002) \\
  26 & (-5.889102546322750E-002,7.627608399874446E-002) \\
  27 & (-9.712149518138669E-002,1.793695100255052E-002) \\
  28 & (0.101272862135273,3.940173722756957E-002) \\
  29 & (8.351246119258422E-002,-8.055956525511754E-002) \\
  30 & (3.447514504354676E-002,-6.113180059952469E-002) \\
  31 & (9.951265147314473E-002,5.575638197924940E-002) \\
  32 & (-8.560101157107276E-002,4.371001847647850E-002) \\
  33 & (1.790860687993339E-002,-4.609380726768647E-002) \\
  34 & (0.101094129379701,6.494214772295025E-002) \\
  35 & (-2.247063699015752E-002,4.864367215816477E-003) \\
  36 & (-0.101021865482900,-3.782742816016475E-002) \\
  37 & (3.152510928837363E-002,0.122475737293311) \\
  38 & (3.278246037718845E-002,-1.256558150969285E-004) \\
  39 & (-5.736492004809834E-002,6.977684817377462E-002) \\
  40 & (2.216141448231444E-002,-7.601939988222593E-002) \\
  41 & (0.131970698296467,-1.260154440769711E-002) \\
  42 & (8.044458687238869E-003,-9.387152676075274E-002) \\
  43 & (-7.808462265554876E-003,-1.202931445781517E-002) \\
  44 & (-3.274238472614039E-002,-2.514421762607319E-002) \\
  45 & (-7.505399199689463E-003,-3.929813385495669E-002) \\
  46 & (0.155137227199514,1.049705149755480E-002) \\
  47 & (3.965712582027887E-002,1.083231718050668E-002) \\
  48 & (-8.224544805028827E-002,-3.383505686446630E-002) \\
  49 & (-0.154734489832632,8.673238144109774E-002) \\
  50 & (-7.332128812157200E-002,-1.371022464291685E-002) \\
  51 & (5.208789441301026E-003,-1.411983814527247E-002) \\
  52 & (-3.590001918998145E-002,4.647625796299270E-002) \\
  53 & (-8.697459750434891E-004,1.482515294565435E-002) \\
  54 & (1.092140821864845E-002,4.129654472949966E-002) \\
  55 & (7.674494499478537E-002,-5.338559685445066E-002) \\
  56 & (-6.251229029986881E-002,6.425293853541948E-002) \\
  57 & (8.520457184967269E-003,-7.709553490818186E-003) \\
  58 & (-3.438221523644015E-002,-9.255954127990704E-002) \\
  59 & (-2.577383579159245E-002,0.129459058820970) \\
  60 & (0.108622543447635,-8.806418991079722E-002) \\
  61 & (-8.106072511646092E-003,3.606461883196400E-002) \\
  62 & (-1.202677529398651E-002,3.058305163904075E-002) \\
  63 & (-2.485595158034444E-002,9.667248785955586E-002) \\
  64 & (6.171068243552971E-002,-9.583626876325756E-003) \\
  65 & (8.806183494115266E-002,-3.526345160182855E-002) \\
  66 & (6.854736532168551E-002,-6.411781011736128E-002)\\
  67 & (2.066804256769957E-002,1.612535204191288E-002)\\
  68 & (1.438805006820953E-002,0.124162489557811)\\
  69 & (-5.074891074532802E-002,-5.439956049423335E-002)\\
  70 & (-3.640086957084941E-002,4.594300372342439E-003)\\
  71 & (3.550293356508465E-002,8.695740710560376E-002)\\
  72 & (-4.773739666022134E-002,-3.667942618866395E-002)\\
  73 & (2.346579563123868E-003,-0.119908858816339)\\
  74 & (1.493075601025749E-002,4.553124163243615E-002)\\
  75 & (5.034836527591473E-002,8.124581001062543E-002)\\
  76 & (6.802270653015219E-002,8.317313465161994E-003)\\
  77 & (6.283616184316396E-002,6.514992784328244E-004)\\
  78 & (0.127829889515795,0.118971821010114)\\
  79 & (-9.788293784579458E-002,5.354297473450592E-003)\\
  80 & (0.117110474490768,-4.317232032001831E-002)\\
  81 & (-9.256055710305476E-002,-2.768362340687266E-002)\\
  82 & (-7.244569839572039E-002,6.671389393930190E-002)\\
  83 & (-5.515716658607148E-002,2.093262220899585E-002)\\
  84 & (-3.028765985082235E-002,4.529684133342195E-002)\\
  85 & (-1.454140943294647E-002,7.974409510449305E-002)\\
  86 & (-7.121856602606923E-002,-4.438866940874264E-002)\\
  87 & (-3.590040749082390E-002,8.143026671780049E-002)\\
  88 & (8.912049927583944E-003,-1.389907243324935E-002)\\
  89 & (9.484845129641119E-002,-5.878664094021236E-002)\\
  90 & (-5.450397076610332E-002,0.117961375334513)\\
  91 & (-1.169436871304801E-002,-6.947913611647639E-002)\\
  92 & (-6.798510500616832E-002,-7.747559839783932E-002)\\
  93 & (1.740724913960769E-002,-1.809038449399666E-002)\\
  94 & (-1.885142661877520E-002,6.314493850061739E-002)\\
  95 & (7.520470652239290E-002,4.456457191590223E-002)\\
  96 & (0.117132792695098,3.066328283226673E-002)\\
  97 & (1.127320363030642E-002,-2.083667932069934E-002)\\
  98 & (1.977443152287268E-002,4.839368466995119E-002)\\
  99 & (-0.146648569587175,-1.841910055111614E-002)\\
  100 &(2.485199104080963E-002,-9.065577146599127E-002) \\
  101 &(-1.352964224869225E-002,-8.518961930320970E-002) \\
  102 &(4.288496230633006E-002,7.033803797783106E-003) \\
  103 &(4.876461334642698E-002,-1.428437645902438E-002) \\
  104 &(-3.244529712612734E-002,8.121540837139055E-002) \\
  105 &(2.809280188171577E-002,4.286033253289921E-002) \\
  106 &(5.009488734831499E-002,-6.852953802160539E-002) \\
  107 &(-4.883631054660045E-002,6.372960434850038E-002) \\
  108 &(-1.583821551197247E-002,-4.855360397290493E-002) \\
  109 &(3.537174285397322E-002,0.104311697071161) \\
  110 &(4.234833191138120E-002,1.152575018630899E-002) \\
  111 &(0.149915848699035,2.063573734200513E-003) \\
  112 &(-2.681850738901102E-003,-2.650438998719609E-002) \\
  113 &(2.099859642637032E-002,7.483425704168839E-002) \\
  114 &(-2.307627608049840E-002,8.294414552141494E-003) \\
  115 &(-7.879700573926614E-002,-5.952656546473500E-002) \\
  116 &(3.702914401846596E-002,5.284665497817300E-003) \\
  117 &(-4.628839981989381E-002,7.345123474109293E-002) \\
  118 &(0.107904736635145,-0.164393350587244) \\
  119 &(4.763528675022823E-002,1.908136182097281E-002) \\
  120 &(0.116908223755807,-4.314878373454251E-002) \\
  121 &(3.495914043033557E-002,-4.526014514286658E-002) \\
  122 &(6.120391755562234E-002,-3.887547264821206E-002) \\
  123 &(3.457915304142278E-002,-7.568701576399368E-002) \\
  124 &(6.046688922765979E-002,-3.864792846188141E-002) \\
  125 &(-3.215267435226381E-002,0.128788000228012) \\
  126 &(-1.191016945303225E-002,3.655884472429104E-003) \\
  127 &(-2.612694626117723E-005,-5.303000737423087E-002) \\
  \hline
\end{longtable}
\end{center}

\end{document}